\documentclass[twocolumn]{aastex7}

\usepackage{graphicx}
\usepackage{amsmath}
\usepackage{hyperref}
\usepackage{longtable}
\usepackage{booktabs}

\newcommand{\te}{t_{\rm E}}
\newcommand{\thetae}{\theta_{\rm E}}

\newcommand{\dl}{D_{\rm L}}
\newcommand{\ds}{D_{\rm S}}
\def\e{{\rm E}}

\definecolor{brown}{rgb}{0.59, 0.29, 0.0}
\definecolor{darkgreen}{rgb}{0.0, 0.42, 0.24}
\definecolor{darkblue}{rgb}{0.01, 0.31, 0.59}
\definecolor{darkblue}{rgb}{0.0, 0.25, 0.42}
\definecolor{blue}{rgb}{0.0,0.0,1.0}
\definecolor{green}{rgb}{0.0,1.0,0.0}

\begin{document}

\title{Four Cold Super-Jupiters Revealed by Extended and Complex Microlensing Signals}
\shorttitle{Four Cold Super-Jupiters}


\author{Cheongho Han}
\affiliation{Department of Physics, Chungbuk National University, Cheongju 28644, Republic of Korea}
\email{cheongho@astroph.chungbuk.ac.kr}
\author{Chung-Uk Lee}
\affiliation{Korea Astronomy and Space Science Institute, Daejon 34055, Republic of Korea}
\email{leecu@kasi.re.kr}
\author{Michael D. Albrow}   
\affiliation{University of Canterbury, Department of Physics and Astronomy, Private Bag 4800, Christchurch 8020, New Zealand}
\email{michael.albrow@canterbury.ac.nz}
\author{Sun-Ju Chung}
\affiliation{Korea Astronomy and Space Science Institute, Daejon 34055, Republic of Korea}
\email{sjchung@kasi.re.kr}
\author{Andrew Gould}
\affiliation{Department of Astronomy, Ohio State University, 140 West 18th Ave., Columbus, OH 43210, USA}
\email{gould.34@osu.edu}
\author{Youn Kil Jung}
\affiliation{Korea Astronomy and Space Science Institute, Daejon 34055, Republic of Korea}
\affiliation{University of Science and Technology, Daejeon 34113, Republic of Korea}
\email{younkil21@gmail.com}
\author{Kyu-Ha~Hwang}
\affiliation{Korea Astronomy and Space Science Institute, Daejon 34055, Republic of Korea}
\email{kyuha@kasi.re.kr}
\author{Yoon-Hyun Ryu}
\affiliation{Korea Astronomy and Space Science Institute, Daejon 34055, Republic of Korea}
\email{yhryu@kasi.re.kr}
\author{Yossi Shvartzvald}
\affiliation{Department of Particle Physics and Astrophysics, Weizmann Institute of Science, Rehovot 76100, Israel}
\email{yossishv@gmail.com}
\author{In-Gu Shin}
\affiliation{Department of Astronomy, Westlake University, Hangzhou 310030, Zhejiang Province, China}
\email{ingushin@gmail.com}
\author{Jennifer C. Yee}
\affiliation{Center for Astrophysics $|$ Harvard \& Smithsonian 60 Garden St., Cambridge, MA 02138, USA}
\email{jyee@cfa.harvard.edu}
\author{Weicheng Zang}
\affiliation{Department of Astronomy, Westlake University, Hangzhou 310030, Zhejiang Province, China}
\email{zangweicheng@westlake.edu.cn}
\author{Hongjing Yang}
\affiliation{Department of Astronomy, Westlake University, Hangzhou 310030, Zhejiang Province, China}
\email{yanghongjing@westlake.edu.cn}
\author{Doeon Kim}
\affiliation{Department of Physics, Chungbuk National University, Cheongju 28644, Republic of Korea}
\email{qso21@hanmail.net}
\author{Dong-Jin Kim}
\affiliation{Korea Astronomy and Space Science Institute, Daejon 34055, Republic of Korea}
\email{keaton03@kasi.re.kr}
\author{Sang-Mok Cha}
\affiliation{Korea Astronomy and Space Science Institute, Daejon 34055, Republic of Korea}
\affiliation{School of Space Research, Kyung Hee University, Yongin, Kyeonggi 17104, Republic of Korea}
\email{chasm@kasi.re.kr}
\author{Seung-Lee Kim}
\affiliation{Korea Astronomy and Space Science Institute, Daejon 34055, Republic of Korea}
\email{slkim@kasi.re.kr}
\author{Dong-Joo Lee}
\affiliation{Korea Astronomy and Space Science Institute, Daejon 34055, Republic of Korea}
\email{marin678@kasi.re.kr}
\author{Yongseok Lee}
\affiliation{Korea Astronomy and Space Science Institute, Daejon 34055, Republic of Korea}
\affiliation{School of Space Research, Kyung Hee University, Yongin, Kyeonggi 17104, Republic of Korea}
\email{yslee@kasi.re.kr}
\author{Byeong-Gon Park}
\affiliation{Korea Astronomy and Space Science Institute, Daejon 34055, Republic of Korea}
\email{bgpark@kasi.re.kr}
\author{Richard W. Pogge}
\affiliation{Department of Astronomy, Ohio State University, 140 West 18th Ave., Columbus, OH 43210, USA}
\email{pogge.1@osu.edu}
\collaboration{100}{(KMTNet Collaboration)}
\correspondingauthor{\texttt{cheongho@astroph.chungbuk.ac.kr}}
\correspondingauthor{\texttt{leecu@kasi.re.kr}}

\begin{abstract}
We present the analysis of four microlensing events, KMT-2020-BLG-0202,
KMT-2022-BLG-1551, KMT-2023-BLG-0466, and KMT-2025-BLG-0121, which exhibit
extended and complex anomalies in their light curves. These events were identified through
a systematic reanalysis of KMTNet data aimed at detecting planetary signals that deviate
from the typical short-term anomaly morphology. Detailed modeling indicates that all four
anomalies were produced by planetary companions to low-mass stellar hosts. The events
have mass ratios of $q \sim (5$–$14)\times10^{-3}$ and Einstein timescales of $t_{\rm E}
\sim 20$–$43$ days. Bayesian analyses based on Galactic models show that the companions
are super-Jupiters with masses of a few to approximately 10 $M_{\rm J}$, orbiting
sub-solar-mass hosts located at distances of $D_{\rm L} \sim 4$--$7$~kpc. All planets lie
well beyond the snow line of their hosts, placing them in the regime of cold giant planets.
These detections demonstrate that extended and complex microlensing anomalies, which are 
often challenging to recognize as planetary in origin, can nonetheless contain planetary 
signals.  This work underscores the unique sensitivity of microlensing to cold, massive 
planets beyond the snow line and highlights the importance of systematic reanalyses of 
survey data for achieving a more complete and unbiased census of exoplanets in the Galaxy.
\end{abstract}

\keywords{\uat{Gravitational microlensing exoplanet detection}{2147}}

\begin{deluxetable*}{lllllll}
\tablewidth{0pt}
\tablecaption{Event coordinates, fields, observational cadence, and $I$-band extinction.  \label{table:one}}
\tablehead{
\multicolumn{1}{c}{ID reference}              &
\multicolumn{1}{c}{(RA, DEC)$_{\rm J2000}$}   &
\multicolumn{1}{c}{$(l, b)$}                  &
\multicolumn{1}{c}{Field}                     &
\multicolumn{1}{c}{Cadence}                   &
\multicolumn{1}{c}{$A_I$}                     
}
\startdata
 KMT-2020-BLG-0202  &  (17:48:09.71, -24:28:28.02)   &  (4$^\circ\hskip-2pt$.1100, 1$^\circ\hskip-2pt$.8282)     &  BLG19  &  1.0 hour  &  2.82  \\
 KMT-2022-BLG-1551  &  (17:41:00.90, -27:33:06.77)   &  (0$^\circ\hskip-2pt$.6457, 1$^\circ\hskip-2pt$.5899)     &  BLG15  &  1.0 hour  &  3.18  \\
 KMT-2023-BLG-0466  &  (18:02:26.40, -33:00:10.69)   &  (-1$^\circ\hskip-2pt$.6930, -5$^\circ\hskip-2pt$.1640)   &  BLG34  &  2.5 hour  &  0.97  \\
 KMT-2025-BLG-0121  &  (17:48:38.01, -23:59:55.68)   &  (4$^\circ\hskip-2pt$.5733, 1$^\circ\hskip-2pt$.9812)     &  BLG19  &  1.0 hour  &  2.09  
\enddata
\end{deluxetable*}

\section{Introduction} \label{sec:one}

Planetary perturbations in gravitational microlensing events typically manifest as 
short-lived anomalies superimposed on an otherwise smooth and symmetric single-lens 
light curve. These deviations, which generally last from a few hours to a few days, 
arise when the source star passes near a small caustic induced by the planet-host 
system \citep{Mao1991, Gould1992}.  Because these signals exhibit distinctive features 
confined to a small region of the light curve, they can be readily identified by 
high-cadence surveys and have served as the primary channel for microlensing planet 
discoveries to date.

However, planetary signals do not always conform to this typical morphology. Under 
certain lens configurations, planets can produce extended and complex anomalies that 
differ substantially from the typical short-duration perturbation. Such cases arise 
when the source trajectory interacts with both the central and planetary caustics or 
when the planet-host separation approaches the resonant regime, where two or three 
caustics merge into a single enlarged structure. Depending on the source trajectory, 
these resonant or near-resonant configurations can smooth, broaden, or elongate the 
magnification pattern, leading to anomalies with multiple substructures.

A major difficulty is that extended planetary perturbations can closely resemble 
anomalies produced by binary lenses that have comparable component masses.  This 
similarity makes it difficult to distinguish planetary systems from equal-mass 
binaries based solely on light-curve shape.  The problem is further compounded by 
the large number of binary-lens events in survey data, which imposes substantial 
practical and computational burdens when searching for the much smaller planetary 
subset.  Consequently, achieving a complete census of planetary microlensing events 
requires systematic and detailed modeling of a large number of anomalous light 
curves.

To address this challenge, \citet[][hereafter Paper~I]{Han2021} carried out a 
dedicated reanalysis of the 2017--2019 microlensing survey data, focusing on planetary 
events whose signatures were difficult to identify because they deviated from the 
standard short-term anomaly morphology.  The analysis begins by identifying candidate 
anomalous events. Each light curve is fit with a 1L1S model, and both cumulative and 
localized residuals are examined to flag statistically significant and temporally 
coherent deviations. Events that meet these criteria are then prioritized for more 
sophisticated modeling, such as planetary, binary-lens, or binary-source interpretations.  
The overall modeling strategy is hierarchical and iterative, advancing from simple 
to increasingly complex descriptions until the observed anomalies are reproduced in 
a physically consistent manner.  This effort uncovered three previously unrecognized 
planets--KMT-2017-BLG-2509Lb, OGLE-2017-BLG-1099Lb, and OGLE-2019-BLG-0299Lb--whose 
light curves displayed extended anomalies produced by source crossings of resonant 
caustics generated by giant planets located near the Einstein rings of their hosts. 
These discoveries demonstrated that planets exhibiting extended and complex anomaly 
patterns can be missed, underscoring the importance of systematic reexaminations of 
microlensing data to improve detection completeness.

Building on the approach of Paper I, we extended this effort to microlensing events 
detected from 2020 to 2025 with the goal of identifying additional planetary systems 
that exhibit extended or complex anomalies.  Through analyses of anomalous lensing 
events detected from this period, we discovered four new planets: KMT-2020-BLG-0202Lb, 
KMT-2022-BLG-1551Lb, KMT-2023-BLG-0466Lb, and KMT-2025-BLG-0121Lb. In this paper, we 
present detailed analyses of these events and discuss their implications for 
understanding planetary systems that generate extended microlensing anomalies.

\section{Observations and data} \label{sec:two}

The four lensing events were detected exclusively by the Korea Microlensing Telescope 
Network \citep[KMTNet;][]{Kim2016} survey. The survey was established to conduct 
high-cadence, wide-field photometric monitoring of dense stellar fields toward the 
Galactic bulge, with the goal of discovering and characterizing gravitational 
microlensing events, including those produced by exoplanets.

KMTNet consists of three identical 1.6 m telescopes, each equipped with a 4 deg$^2$ 
field-of-view camera, located at the Cerro Tololo Inter-American Observatory in Chile 
(KMTC), the South African Astronomical Observatory in South Africa (KMTS), and the 
Siding Spring Observatory in Australia (KMTA). This global configuration enables 
continuous, round-the-clock monitoring of the Galactic bulge, allowing high-cadence 
observations throughout the bulge season. Most images are obtained in the $I$ band, 
with roughly 10\% $V$-band frames acquired for color information.

Table~\ref{table:one} lists the equatorial and Galactic coordinates of the analyzed events, 
along with the corresponding KMTNet fields, observation cadences, and $I$-band extinctions 
toward the respective fields.  For KMT-2022-BLG-1551, supplementary data were acquired 
through follow-up observations conducted with the 1.0 m telescope of the Las Cumbres 
Observatory at the South African Astronomical Observatory site (LCOS). The follow-up 
images were acquired at two main epochs: near the peak of the event and during the 
declining phase of the light curve. All follow-up observations were conducted in the 
$I$ band.

\begin{deluxetable}{llrllll}
\tablewidth{0pt}
\tablecaption{Number of data points and error-bar readjustment factors.  \label{table:two}}
\tablehead{
\multicolumn{1}{c}{Telescope }            &
\multicolumn{1}{c}{Data set }             &
\multicolumn{1}{c}{$N_{\rm data}$ }       &
\multicolumn{1}{c}{$k$ }                  &
\multicolumn{1}{c}{$\sigma_{\rm min}$ }   
}
\startdata
 KMT-2020-BLG-0202  & KMTA19   & 441   &  0.843  & 0.010   \\
\hline
 KMT-2022-BLG-1551  & KMTC15   & 945   &  1.217  & 0.020   \\
                    & KMTS15   & 583   &  1.288  & 0.020   \\
                    & KMTA15   & 295   &  1.513  & 0.010   \\
                    & LCOS     &  32   &  0.949  & 0.004   \\
\hline
 KMT-2023-BLG-0466  & KMTC34   & 493   &  0.648  & 0.010   \\
                    & KMTS34   & 237   &  0.980  & 0.010   \\
                    & KMTA34   & 266   &  0.794  & 0.010   \\
\hline
 KMT-2025-BLG-0121  & KMTC19   & 1012  &  0.806  & 0.020   \\
                    & KMTS19   & 438   &  0.787  & 0.020   \\
                    & KMTA19   & 425   &  0.830  & 0.020   
\enddata
\end{deluxetable}

Both the survey and follow-up data were reduced with the KMTNet photometric pipeline 
\citep{Albrow2009}, which is based on the difference-image-analysis (DIA) method 
\citep{Tomaney1996, Alard1998, Wozniak2000}.  This technique is particularly effective 
for crowded stellar fields toward the Galactic bulge, where blending is severe, and it 
yields high-precision relative photometry suitable for microlensing analyses.  For 
optimal data fidelity, the KMTNet dataset was reprocessed with the new tender-love 
care photometry pipeline described in \citet{Yang2024}.   For each event, the photometry 
from the three KMTNet sites was processed independently and then combined to construct 
its light curve.  During this process, the photometric uncertainties were empirically 
renormalized so that each data set satisfies $\chi^2/{\rm dof}\simeq1$ for the best-fit 
model. The renormalization followed the procedure described by \citet{Yee2012}.  In 
particular, the error bars were adjusted according to
\begin{equation}
\sigma' = k \sqrt{\sigma^2 + \sigma_{\rm min}^2},
\label{eq1}
\end{equation}
\hskip-4pt
where $\sigma$ denotes the reported uncertainty, $k$ is a multiplicative scale factor, 
and $\sigma_{\rm min}$ represents a systematic error floor.  In Table~\ref{table:two}, 
we list the error-bar renormalization factors and the number of data points ($N_{\rm data}$) 
for each data set.

\begin{figure*}[t]
\centering
\includegraphics[width=11.0cm]{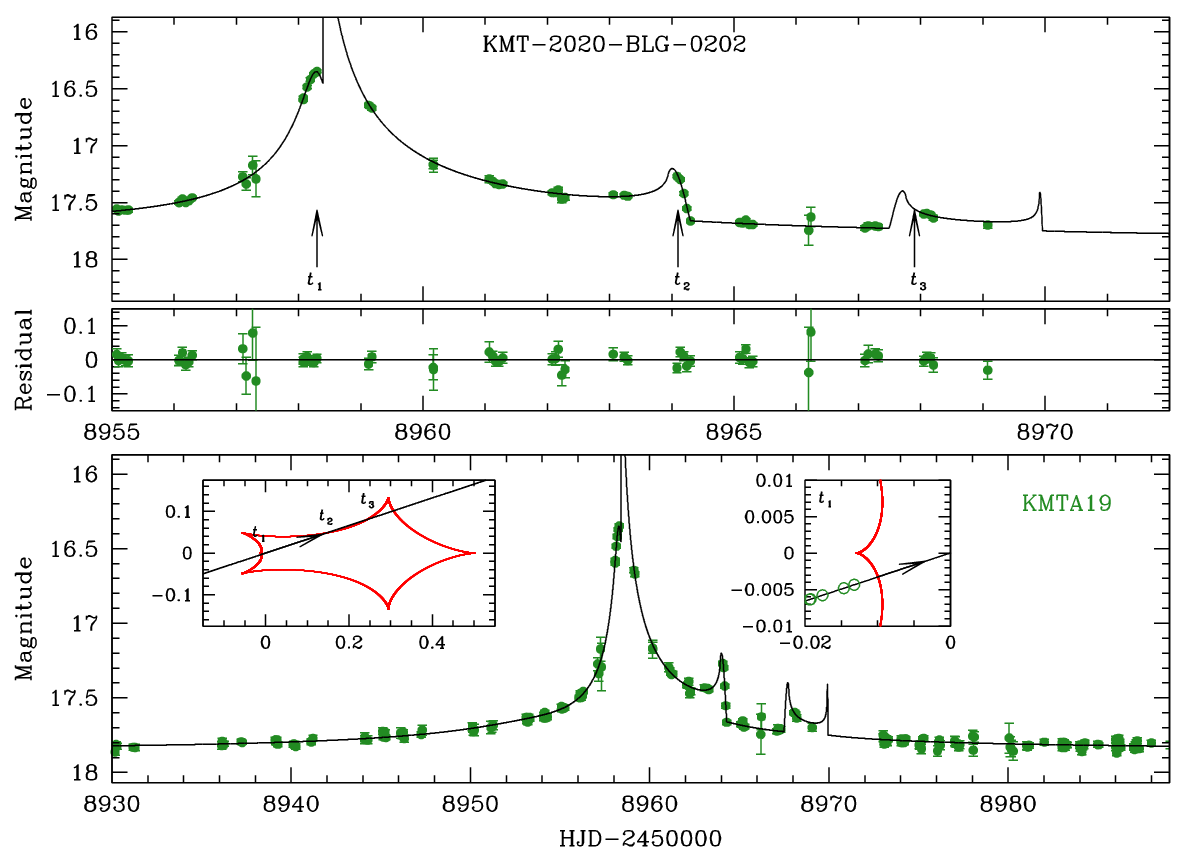}
\caption{
Light curve of the microlensing event KMT-2020-BLG-0202. The bottom panel presents
the overall light curve, while the two upper panels show the zoomed-in view of the 
anomaly region and the residuals from the best-fit model (solid curve). The arrows 
labeled $t_1$, $t_2$, and $t_3$ in the top panel mark the times of the major anomalies. 
The left inset in the bottom panel illustrates the lens-system geometry, depicting 
the source trajectory (arrowed line) with respect to the caustic structure (red 
cuspy curve) generated by the binary lens. The source locations at the epochs of 
the major anomalies are indicated by labels matching those in the top panel. 
The right inset provides a magnified view of the configuration in the vicinity of 
$t_1$.  The small green circles along the source trajectory mark the source positions 
at the epochs of data acquisition. Their sizes are scaled to the actual source size. 
}
\label{fig:one}
\end{figure*}

\begin{figure}[t]
\includegraphics[width=\columnwidth]{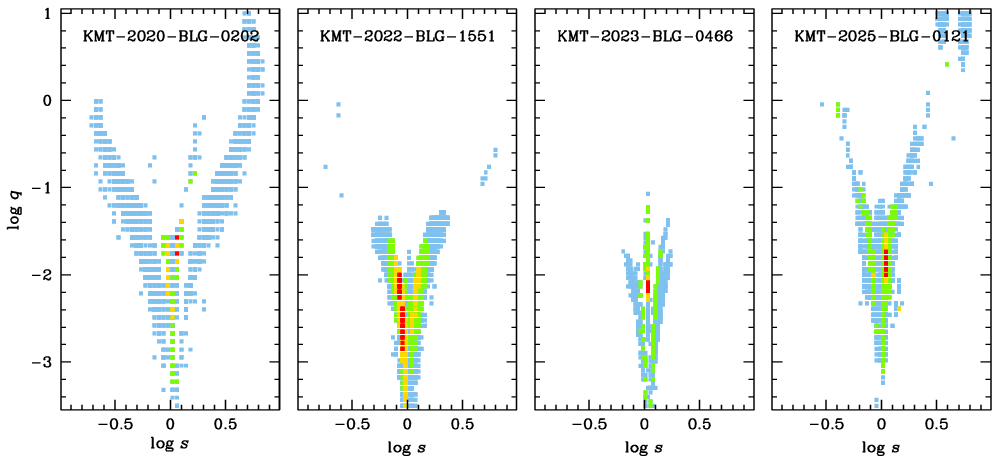}
\caption{
$\Delta\chi^2$ maps in the $(s,q)$ grid-parameter space. The color scale indicates 
regions with $\Delta\chi^2 \leq n\times 1^2$ (red), $n\times 2^2$ (yellow), $n\times 
3^2$ (green), and $n\times 4^2$ (cyan), where $n=10$ for KMT-2020-BLG-0202 and 
KMT-2022-BLG-1551, $n=4$ for KMT-2023-BLG-0466, and $n=15$ for KMT-2025-BLG-0121.
}
\label{fig:two}
\end{figure}

\section{Light curve analyses} \label{sec:three}

The light curves of all analyzed events show prominent caustic-related features, 
indicating that the lenses are multiple systems. Accordingly, we begin our 
analysis by modeling the light curves with a binary-lens single-source (2L1S) framework.

The modeling of the light curve was carried out to determine a lensing solution, which 
represents a set of parameters describing the observed light curve. The light curve of 
a 2L1S event is characterized by seven basic parameters $(t_0, u_0, \te, s, q, \alpha, 
\rho)$.  Three of these describe the lens-source geometry: $t_0$ is the time of the 
closest approach between the source and the reference position of the lens, $u_0$ is 
the impact parameter (the lens–source separation at $t_0$) in units of the angular 
Einstein radius $\thetae$, and $\te$ is the Einstein timescale, defined as the time 
required for the source to traverse $\thetae$. Two additional parameters characterize 
the binary nature of the lens: $s$ denotes the projected separation between the lens 
components ($M_1$ and $M_2$) in units of $\thetae$, and $q$ represents their mass ratio. 
The parameter $\alpha$ specifies the orientation of the source trajectory with respect 
to the binary-lens axis. Finally, the parameter $\rho$, defined as the ratio of the 
angular source radius to $\thetae$, accounts for finite-source effects that become 
important during caustic crossings or approaches.

We search for a lensing solution in two stages. We first perform a grid search over $(s,q)$, 
for which the magnification is highly sensitive to small parameter changes, while optimizing 
the remaining parameters via downhill minimization with initial guesses based on basic 
light-curve features such as event epoch, peak magnification, and timescale. For the 
source-trajectory angle $\alpha$, we use 20 uniformly spaced trial values over $0$--$2\pi$. 
The resulting $\chi^2$ maps in the $(s,q)$ plane are used to locate local minima (candidate 
solutions). We then refine each candidate by allowing all parameters to vary simultaneously. 
If multiple solutions remain, we assess degeneracies by comparing their $\chi^2$ values. 
All modeling is carried out with a custom code based on the map-making method of 
\citet{Dong2006}, using ray shooting to compute finite-source magnifications.  In our 
modeling, we do not include higher-order effects because the event timescales are not 
long enough and the light curves show no clear signatures of such effects.

In the following subsections, we present detailed analyses of the individual events. For each 
event, we begin by briefly describing the data sets used and identifying the distinctive anomaly 
observed in the light curve. We then present the corresponding lensing model and best-fit solution, 
followed by a discussion of how the observed anomaly is explained within the binary-lens framework.

\begin{figure*}[t]
\centering
\includegraphics[width=11.0cm]{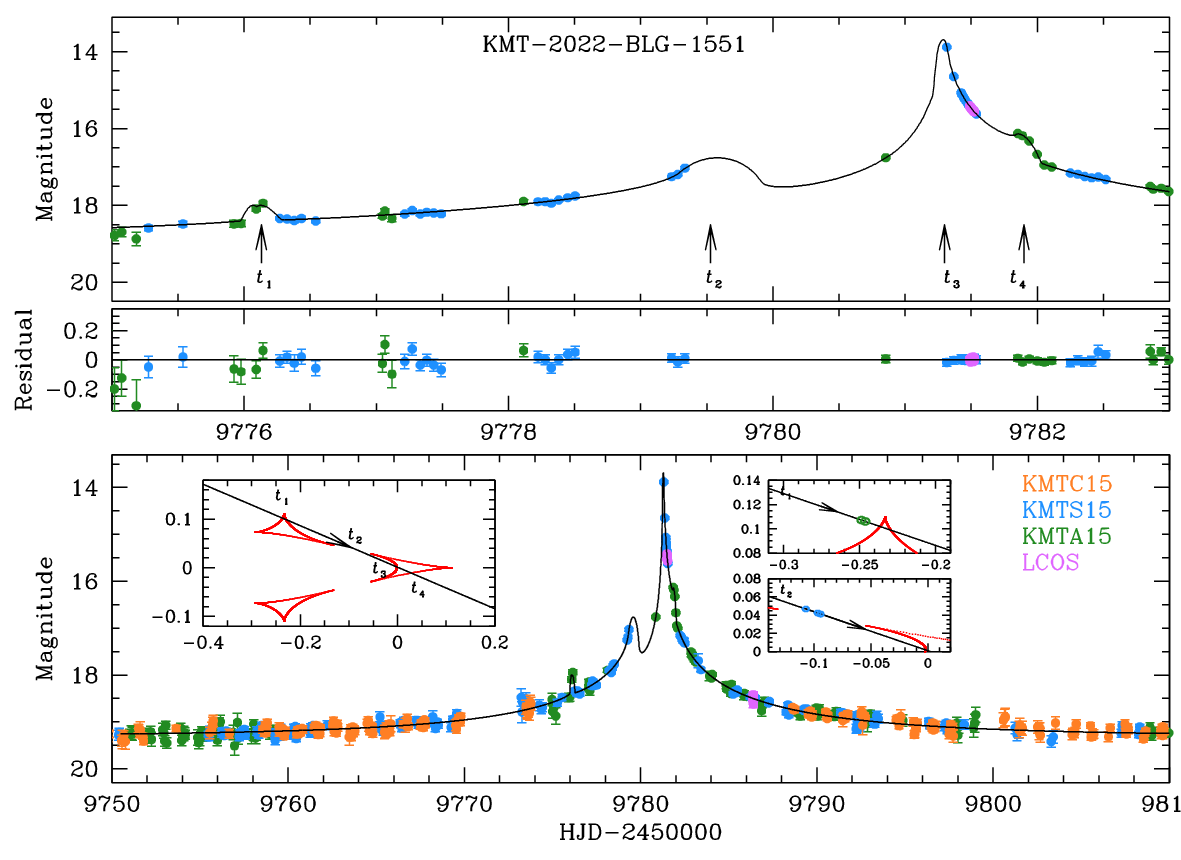}
\caption{
Lensing light curve of KMT-2022-BLG-1551.  Notations are consistent with those used in 
Fig.~\ref{fig:one}.  The two right insets in the bottom panel show zoomed-in views of the 
lens-system configurations around $t_1$ and $t_2$, respectively.  The small circles along 
the source trajectory mark the source positions near the time of the anomaly, with their 
sizes scaled to the source radius and their colors matching those of the telescopes shown 
in the legend.
}
\label{fig:three}
\end{figure*}

\subsection{KMT-2020-BLG-0202} \label{sec:three-one}

The lensing event KMT-2020-BLG-0202 was detected on 2020 April 17, which corresponds to the 
abridged Heliocentric Julian date ${\rm HJD}^\prime \equiv {\rm HJD} - 2450000 = 8956$. The 
baseline magnitude of the source is $I_{\rm base}=17.84$.  During most of the 2020 season, 
two KMTNet telescopes, KMTC and KMTS, were shut down due to the COVID-19 pandemic. Consequently, 
the event was observed solely by the KMTA telescope, which remained operational throughout the 
2020 season.

Figure~\ref{fig:one} shows the lensing light curve of the event.  Despite the relatively sparse 
coverage due to single-telescope observations, the light curve exhibits an extended complex 
anomaly pattern comprising three major features around $t_1 = 8958.3$, $t_2 = 8964.1$, and $t_3 
= 8967.9$. From the sharp rises and falls, it is likely that these features result from caustic 
crossings of the source star. Based on the pattern, the two features at $t_1$ and $t_2$ appear 
to form a pair corresponding to the source’s caustic entrance and exit. The feature at $t_3$ 
also appears to result from a caustic entrance, although the corresponding exit was not covered. 
The anomalous region constitutes an important portion of the light curve, and it was therefore 
initially suspected that the anomaly was produced by a binary lens composed of two components 
with roughly equal masses.

Despite the significant deviation from the typical short-duration anomaly pattern, the 
anomaly was determined to be of planetary origin.  Modeling of the light curve yielded a 
unique solution with binary parameters of $(s, q) \sim (1.16, 14.4 \times 10^{-3})$.  The 
left panel of Figure~\ref{fig:two} presents the corresponding $\Delta\chi^2$ map in the 
$(s,q)$ plane. We note that similarly unique solutions are obtained not only for 
KMT-2020-BLG-0202 but also for the other events, as shown in the remaining maps.  The 
complete set of lensing parameters for the event is presented in Table~\ref{table:three}, 
and the model light curve is shown as a solid line in Figure~\ref{fig:one}. The mass ratio 
is approximately 15 times greater than that between Jupiter and the Sun. Considering that 
Galactic microlensing events are typically produced by low-mass stellar hosts \citep{Han2003}, 
the companion mass is likely below the upper planetary limit of $\sim 13~M_{\rm J}$. The 
event timescale is estimated to be $t_{\rm E} \sim 35$~days. From the resolved caustic 
feature around $t_2$, the normalized source radius was measured to be $\rho \sim 0.74 
\times 10^{-3}$.

\begin{deluxetable}{lllllll}
\tablewidth{0pt} 
\tablecaption{Lensing parameters of KMT-2020-BLG-0202 and KMT-2022-BLG-1551.\label{table:three}}
\tablehead{
\multicolumn{1}{c}{Parameter}                &
\multicolumn{1}{c}{KMT-2020-BLG-0202}        &
\multicolumn{1}{c}{KMT-2022-BLG-1551}        
}
\startdata
 $\chi^2$             &  $427.0             $    &  $1858.9              $  \\  
 $t_0$ (HJD$^\prime$) &  $8958.779 \pm 0.031$    &  $9781.3238 \pm 0.0027$  \\
 $u_0$ ($10^{-3}$)    &  $0.031 \pm 0.621   $    &  $0.71 \pm 0.29       $  \\
 $\te$ (days)         &  $34.50 \pm 1.50    $    &  $19.31 \pm 0.65      $  \\
 $s$                  &  $1.1577 \pm 0.0059 $    &  $0.8840 \pm 0.0027   $  \\
 $q$ ($10^{-3}$)      &  $14.39 \pm 1.79    $    &  $5.26 \pm 0.57       $  \\
 $\alpha$ (rad)       &  $2.822 \pm 0.016   $    &  $3.581 \pm 0.022     $  \\
 $\rho$ ($10^{-3}$)   &  $0.736 \pm 0.068   $    &  $3.13 \pm 0.29       $  \\
\enddata
\tablecomments{ ${\rm HJD}^\prime \equiv {\rm HJD} - 2450000$.  }
\end{deluxetable}

The configuration of the lens system is shown in the inset of the bottom panel of 
Figure~\ref{fig:one}. The caustic induced by the binary lens forms a single resonant 
structure comprised of six folds meeting at six cusps. The source traversed the caustic 
from the lower-left to the upper-right.  It initially passed near the left on-axis cusp 
and subsequently crossed the lower-left fold. These interactions generated a complex 
light-curve morphology, including a weak bump at ${\rm HJD}^\prime \sim 8958.3$ 
and a sharp caustic spike at ${\rm HJD}^\prime \sim 8958.6$.  Owing to the concavity 
of the caustic fold, the source reentered the caustic by passing through a different 
portion of the upper fold and finally exited through the upper-right fold. The first 
caustic passage of the source corresponds to the $t_1$--$t_2$ caustic pair, while the 
second passage corresponds to another caustic pair with the entrance at $t_3$. The 
light-curve pattern between $t_1$ and $t_2$ deviates from a typical U-shaped profile 
because the source approached the upper fold asymptotically while moving inside the 
caustic.

\begin{figure*}[t]
\centering
\includegraphics[width=11.0cm]{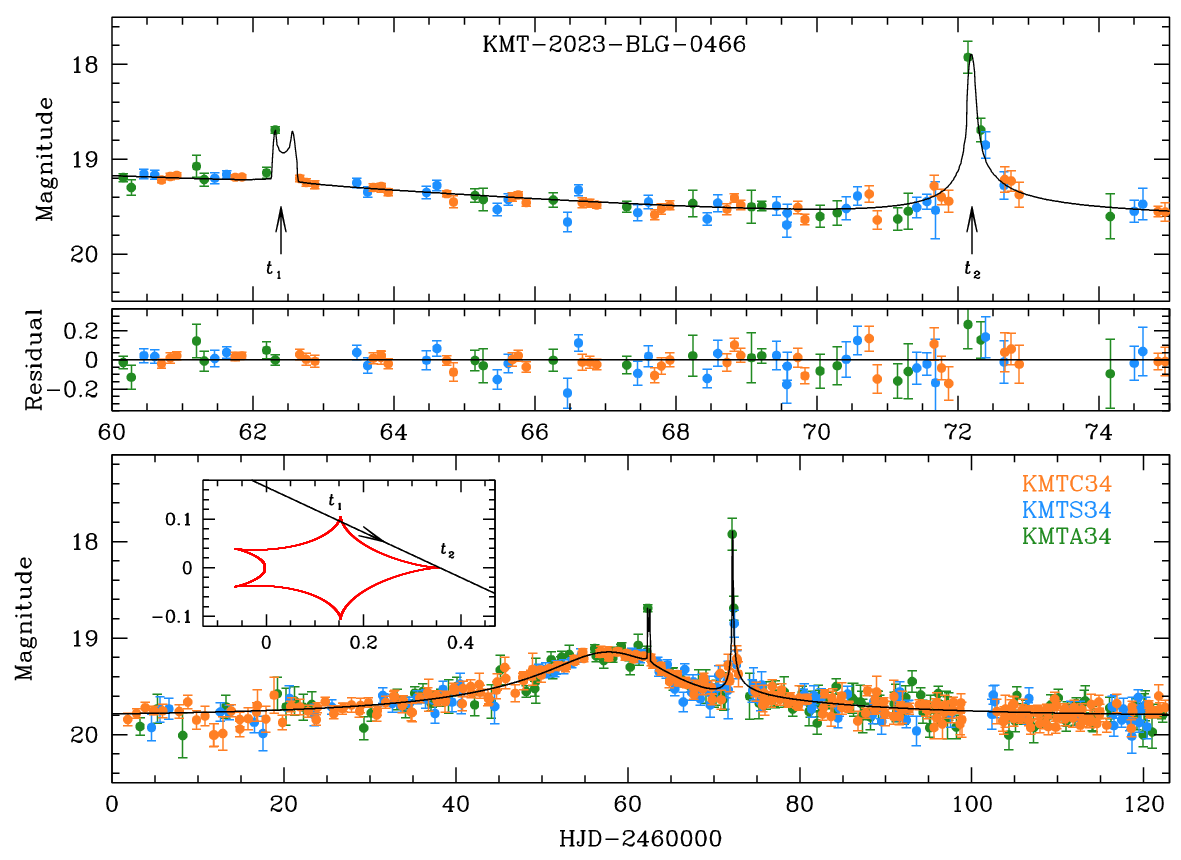}
\caption{
Light curve of the lensing event KMT-2023-BLG-0466.
The notation is the same as in Figure~\ref{fig:one}.
}
\label{fig:four}
\end{figure*}

\subsection{KMT-2022-BLG-1551} \label{sec:three-two}

The lensing event KMT-2022-BLG-1551 occurred on a source with a baseline magnitude of
$I_{\rm base} = 19.48$ during the 2022 season. It was detected by the KMTNet survey on 
2022 July 19 (${\rm HJD}^\prime = 9779$), approximately two days before the event reached 
a very high magnification of $A_{\rm max} \sim 425$. Because the peak region of a 
high-magnification event is highly sensitive to planetary perturbations \citep{Griest1998}, 
follow-up observations were conducted with the LCOS telescope shortly after the peak.
In addition, KMTS activated its ``auto-followup'' mode, in which the cadence was increased 
to about 3.5/hr, beginning at approximately ${\rm HJD}^\prime \sim 9781.42$ and continuing 
until the end of the night at $\sim 9781.54$.

Figure~\ref{fig:three} shows the lensing light curve of KMT-2022-BLG-1551. The peak region 
exhibits an extended and complex anomaly pattern composed of multiple distinct features. 
The first brief anomaly feature appears at $t_1 \sim 9776.1$, followed by a smoother feature 
at $t_2 \sim 9779.5$. The third and fourth features, forming a caustic pair, occur at $t_3 
\sim 9781.3$ and $t_4 \sim 9781.9$, respectively. No KMTC data were obtained during the 
anomaly because the telescope was under maintenance. The anomaly region spans nearly one-third 
of the total event timescale, deviating from the typical morphology of a planetary perturbation.

Despite the atypical anomaly pattern, modeling of the light curve indicates that the 
perturbations were produced by a planetary companion to the primary lens.  The unique 
binary-lens solution yields parameters $(s,q) \sim (0.88, 5.3\times10^{-3})$ and an event 
timescale of $t_{\rm E} \sim 19$~days.  Analysis of the anomaly features affected by 
finite-source effects provides a normalized source radius of $\rho \sim 3.1 \times 10^{-3}$.  
Table~\ref{table:three} lists the complete set of lensing parameters for the event, and 
Figure~\ref{fig:three} presents the model light curve.

The inset in the bottom panel of Figure~\ref{fig:three} shows the configuration of the lens 
system.  The planet induces two sets of caustics: a central caustic consisting of four 
folds, and two planetary caustics, each composed of three folds located above and below 
the planet--host axis.  The source trajectory passes through both the central and planetary 
caustics. It first crossed the tip of the upper planetary caustic and then moved through the 
region between the central and planetary caustics, producing the successive anomaly features 
at $t_1$ and $t_2$. Subsequently, the source traversed the central caustic, giving rise to 
the pair of caustic-crossing features observed at $t_3$ and $t_4$.

\begin{figure}[t]
\includegraphics[width=\columnwidth]{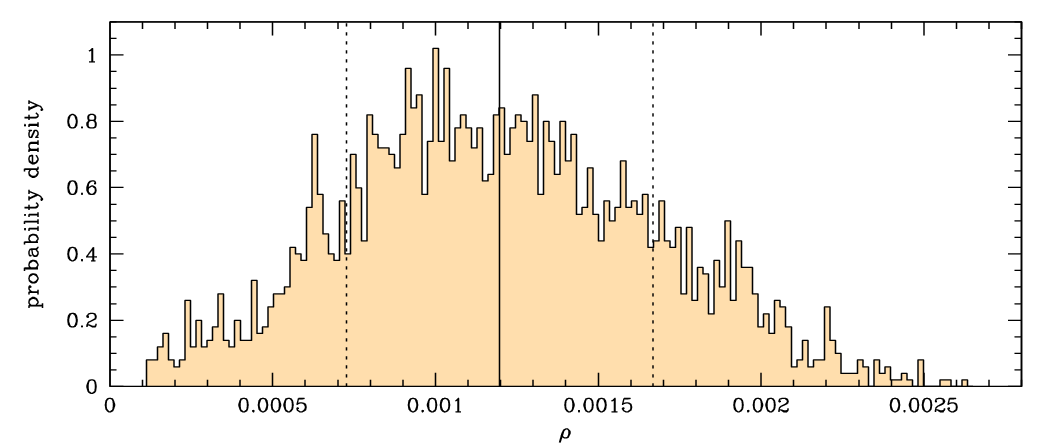}
\caption{
Distribution of the normalized source radius in the MCMC chain for KMT-2023-BLG-0466. The 
vertical solid line denotes the mean value, and the two dotted lines mark the 1$\sigma$ 
interval.
}
\label{fig:five}
\end{figure}

\begin{figure*}[t]
\centering
\includegraphics[width=11.0cm]{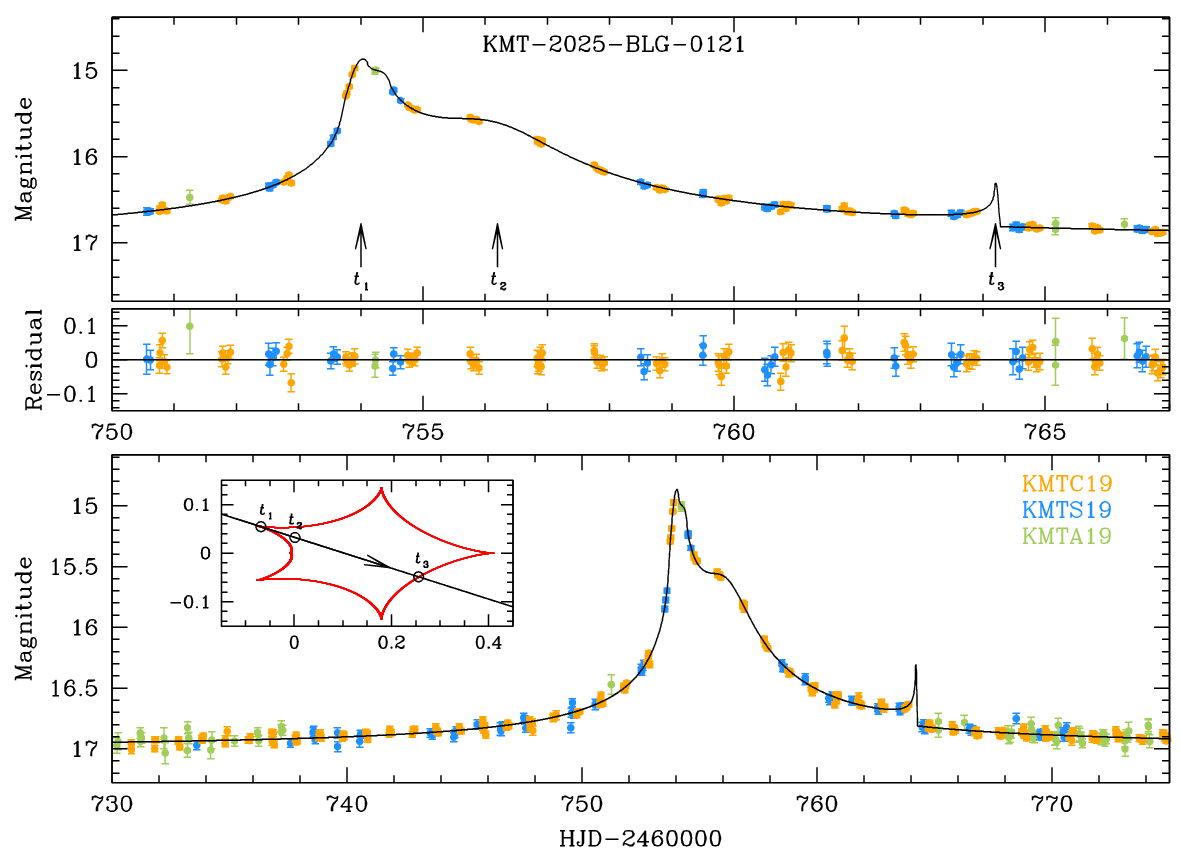}
\caption{
Light curve of the lensing event KMT-2025-BLG-0121.
}
\label{fig:six}
\end{figure*}

\subsection{KMT-2023-BLG-0466} \label{sec:three-three}

The lensing-induced flux magnification of the event KMT-2023-BLG-0466 was first detected
on 2023 April 18 (${\rm HJD}^\prime \equiv {\rm HJD} - 2460000 = 52$), before the light
curve reached a moderate peak magnification of $A_{\rm max} \sim 6.5$. After the peak, the
light curve exhibited multiple anomalies on the declining side.

Figure~\ref{fig:four} shows the lensing light curve of the event.  The first anomaly 
occurred at ${\rm HJD}^\prime \sim 62.4$ ($t_1$) and the second at ${\rm HJD}^\prime \sim 
72.2$ ($t_2$).  Although each feature appears to be a short-term perturbation of a typical 
planetary signal, their successive occurrence with a time interval of 9.8 days makes the 
overall anomaly unusual. While the second anomaly was covered by multiple data points 
from all three KMTNet telescopes, the first was recorded by a single data point from KMTA.  
At first glance, the KMTA data point at $t_1$ appeared to be an outlier, but subsequent 
modeling confirmed that it is a genuine signal. Owing to its large temporal separation 
from the feature at $t_2$, it tightly constrains the source trajectory and, consequently, 
the remaining lensing parameters.

The 2L1S modeling yielded a unique solution with binary parameters of $(s, q) \sim (1.08, 
7.0 \times 10^{-3})$. The low mass ratio indicates that the companion to the lens is 
a planetary object. The event timescale is $\te \sim 43$~days.  To assess whether the 
normalized source radius is actually constrained, we examined its distribution in the 
MCMC chain. As shown in Figure~\ref{fig:five}, the distribution indicates that $\rho$ 
is measured, although its uncertainty remains substantial because the anomaly is only 
partially covered.  Table~\ref{table:four} lists the complete set of lensing parameters 
for the solution, and Figure~\ref{fig:four} shows the corresponding model light curve.

\begin{deluxetable}{lllllll}
\tablewidth{0pt}
\tablecaption{Lensing parameters of KMT-2023-BLG-0466 and KMT-2025-BLG-0121. \label{table:four}}
\tablehead{
\multicolumn{1}{c}{Parameter}                &
\multicolumn{1}{c}{KMT-2023-BLG-0466}        &
\multicolumn{1}{c}{KMT-2025-BLG-0121}        
}
\startdata
 $\chi^2$             &  $998.8            $    &  $1880.2             $  \\  
 $t_0$ (HJD$^\prime$) &  $58.15 \pm 0.14   $    &  $756.463 \pm 0.020  $  \\
 $u_0$ ($10^{-3}$)    &  $0.1505 \pm 0.0082$    &  $0.0311 \pm 0.0010  $  \\
 $\te$ (days)         &  $43.43 \pm 1.68   $    &  $29.97 \pm 0.54     $  \\
 $s$                  &  $1.0779 \pm 0.0037$    &  $1.0912 \pm 0.0026  $  \\
 $q$ ($10^{-3}$)      &  $6.97 \pm 0.89    $    &  $12.1 \pm 0.74      $  \\
 $\alpha$ (rad)       &  $3.577 \pm 0.012  $    &  $3.449 \pm 0.011    $  \\
 $\rho$ ($10^{-3}$)   &  $1.20 \pm 0.47    $    &  $1.13 \pm 0.08      $  
\enddata
\tablecomments{${\rm HJD}^\prime \equiv {\rm HJD} - 2460000$.  }
\end{deluxetable}

The inset in the bottom panel of Figure~\ref{fig:four} shows the lens-system configuration
of the event. The planetary lens produces a resonant caustic similar in shape to that of 
KMT-2020-BLG-0202.  However, the source trajectory relative to the caustic is markedly 
different. Rather than crossing the caustic, the source passes along its outer region, 
grazing it twice--first near the upper cusp and then near the right on-axis cusp. These two 
grazing encounters produce the anomaly features observed at $t_1$ and $t_2$.  Because of the 
large size of the caustic and the spatial separation between the two source--caustic encounters, 
the resulting anomalies appear well separated in time.

\begin{figure*}[t]
\centering
\includegraphics[width=14.5cm]{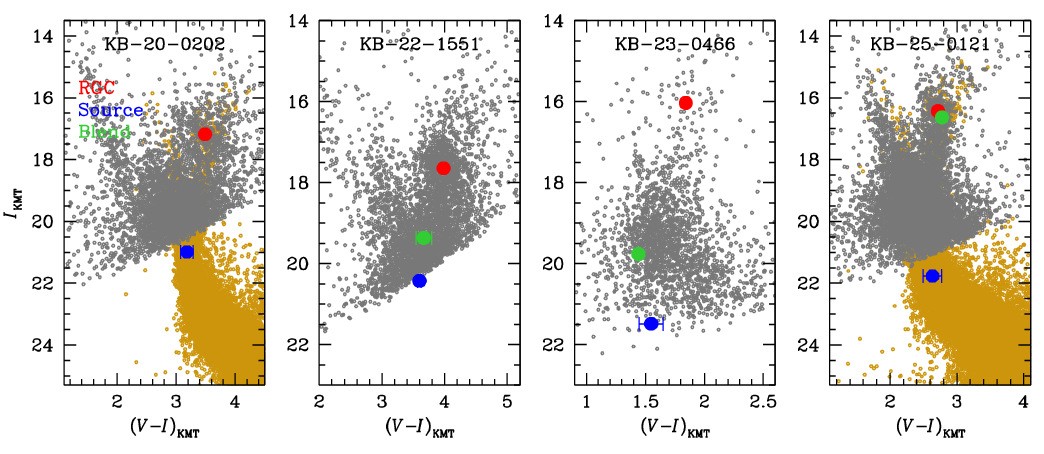}
\caption{
Locations of the source and the centroid of the red giant clump (RGC) in the instrumental
color-magnitude diagrams (CMDs).  For the two events KMT-2020-BLG-0202 and KMT-2025-BLG-0121, 
the source colors were estimated using combined CMDs constructed from KMTC observations 
(gray dots) and HST data (brown dots).  For KMT-2022-BLG-1551, KMT-2023-BLG-0466, and 
KMT-2025-BLG-0121, the positions of the blend stars are also indicated.
}
\label{fig:seven}
\end{figure*}

\subsection{KMT-2025-BLG-0121} \label{sec:three-four}

The lensing event KMT-2025-BLG-0121 occurred on an apparently  bright source with a baseline
magnitude of $I_{\rm base} = 16.97$. The event was detected during its early magnification 
phase on 2025 March 18 (${\rm HJD}^\prime = 752$).

The light curve of the event is presented in Figure~\ref{fig:six}. It exhibits a noticeable 
asymmetry with respect to the peak at ${\rm HJD}^\prime = 754.0$ ($t_1$). The portion of the 
light curve following $t_1$ displays two distinct features: a smooth bump centered at 
${\rm HJD}^\prime = 756.2$ ($t_2$) and a sharp caustic spike at ${\rm HJD}^\prime = 764.2$ 
($t_3$). Together, these features form an extended anomaly that occupies a substantial fraction 
of the light curve, making it difficult to infer its planetary origin based solely on the 
observed light-curve morphology.

Contrary to initial expectations, modeling of the light curve indicates that the anomaly 
was produced by a very low-mass companion to the primary lens.  The light curve is uniquely 
described by a solution with binary parameters $(s, q)\sim (1.09, 12.1\times10^{-3})$ and 
an event timescale of $t_{\rm E}\sim 30$~days.  Table~\ref{table:four} lists the complete 
set of lensing parameters, and the model light curve is shown as a solid line in 
Figure~\ref{fig:six}. The event is found to be heavily blended, indicating that the 
relatively bright baseline flux is dominated by blended light rather than by the source 
itself. The normalized source radius was precisely determined from the well solved anomaly 
feature around $t_1$.

The configuration of the lens system is illustrated in the inset of the lower panel of 
Figure~\ref{fig:six}.  The planetary lens produces a resonant caustic structure similar 
in morphology to those of KMT-2020-BLG-0202 and KMT-2023-BLG-0466. The source traverses 
the caustic diagonally, entering through the upper-left cusp and exiting through the 
lower-right fold. The peak at $t_1$ corresponds to the time when the source entered the 
caustic, while the bump around $t_2$ arises as the source asymptotically approached a 
nearby caustic fold. The feature at $t_3$ corresponds to the time of the caustic exit.

\begin{deluxetable*}{lllllll}
\tablewidth{0pt}
\tablecaption{Source parameters, angular Einstein radius, and relative lens-source proper motion.  \label{table:five}}
\tablehead{
\multicolumn{1}{c}{Parameter}               &
\multicolumn{1}{c}{KMT-2020-BLG-0202}       &
\multicolumn{1}{c}{KMT-2022-BLG-1551}       &
\multicolumn{1}{c}{KMT-2023-BLG-0466}       &
\multicolumn{1}{c}{KMT-2025-BLG-0121}       
}
\startdata
 $(V-I)$                      &   $3.184 \pm 0.106 $   &  $3.599 \pm 0.028 $  &  $1.549 \pm 0.104 $    &   $2.634 \pm 0.139 $    \\
 $I$                          &   $20.985 \pm 0.038$   &  $20.424 \pm 0.001$  &  $21.484 \pm 0.024$    &   $21.772 \pm 0.010$    \\
 $(V-I, I)_{\rm RGC}$         &   $(3.492, 17.179) $   &  $(3.983, 17.646) $  &  $(1.843, 16.034) $    &   $(2.718, 16.430) $    \\
 $(V-I, I)_{{\rm RGC},0}$     &   $(1.060, 14.326) $   &  $(1.060, 14.339) $  &  $(1.060, 14.538) $    &   $(1.060, 14.389) $    \\
 $(V-I)_0$                    &   $0.752 \pm 0.106 $   &  $0.675 \pm 0.049 $  &  $0.766 \pm 0.111 $    &   $0.976 \pm 0.139 $    \\
 $I_0$                        &   $18.133 \pm 0.038$   &  $17.117 \pm 0.020$  &  $19.988 \pm 0.031$    &   $19.731 \pm 0.010$    \\
  Spectral type               &    G8V                 &   G1V (turnoff)      &   G7V                  &    K2.5V                \\
 $\theta_*$ ($\mu$as)         &   $0.780 \pm 0.099 $   &  $1.139 \pm 0.097 $  &  $0.337 \pm 0.044 $    &   $0.499 \pm 0.078 $    \\
 $\thetae$ (mas)              &   $1.060 \pm 0.166 $   &  $0.364 \pm 0.046 $  &  $0.281 \pm 0.116 $    &   $0.442 \pm 0.076 $    \\
 $\mu$ (mas/yr)               &   $11.22 \pm 1.83  $   &  $6.89 \pm 0.8    $  &  $2.36 \pm 0.98   $    &   $5.39 \pm 0.93   $    
\enddata
\end{deluxetable*}

\section{Angular Einstein radius} \label{sec:four}

For a lensing event with a measured normalized source radius, the angular Einstein radius 
can be determined as
\begin{equation}
\theta_{\rm E} = {\theta_* \over \rho}, 
\label{eq1}
\end{equation}
\hskip-3pt
where $\theta_*$ denotes the angular radius of the source. Measuring the angular Einstein 
radius is crucial for constraining the physical lens parameters because it is related to 
the lens mass ($M$) and distance ($\dl$) as
\begin{equation}
\theta_{\rm E} = \sqrt{\kappa M \pi_{\rm rel}}. 
\label{eq2}
\end{equation}
\hskip-4pt
Here $\kappa = 4G/(c^2 {\rm AU})$ and $\pi_{\rm rel} = {\rm AU}(D_{\rm L}^{-1} - D_{\rm
S}^{-1})$ is the relative lens-source parallax, and $\ds$ denotes the distance to the 
source. For all analyzed events, the normalized source radii were measured. Therefore, 
to determine $\theta_{\rm E}$, it is necessary to estimate $\theta_*$.

We estimated the angular radius of the source star from its reddening- and extinction-corrected
(de-reddened) color and magnitude, $(V-I, I)_0$. The procedure was as follows. First, we
measured the instrumental magnitudes of the source in the $V$ and $I$ bands. Second, we 
located the source on the color--magnitude diagram (CMD) constructed from stars in the
vicinity of the event. Third, we calibrated the source's color and magnitude using a reference
feature in the CMD whose de-reddened values are known. Finally, we derived the angular radius 
of the source from an empirical relation between stellar color, magnitude, and angular radius.

We measured the source magnitude by fitting the observed light curve with the model. For this
purpose, we used light curves processed with the pyDIA photometry code \citep{Albrow2017},
which was also used consistently to construct the CMD. For color and magnitude calibration, we
adopted the centroid of the red giant clump (RGC) in the CMD as a reference, whose de-reddened
color and magnitude were taken from previous studies by \citet{Bensby2013} and \citet{Nataf2013}, 
respectively. We then derived the angular source radius using the $(V-K, I)-\theta_*$ relation. 
Because this relation requires the $V-K$ color, we converted the measured $V-I$ color into $V-K$ 
using the color--color relation of \citet{Bessell1988}.

For the two events KMT-2020-BLG-0202 and KMT-2025-BLG-0121, it was difficult to obtain a
reliable source color because of the challenges in measuring the $V$-band magnitude. In these
cases, we derived the source color by first aligning the CMD constructed from KMTC observations
with that from HST data \citep{Holtzman1998}. We then inferred the source color as the mean
value of stars along the main-sequence branch within the range corresponding to the $I$-band
magnitude offset from the RGC centroid.

Figure~\ref{fig:seven} shows the locations of the source stars in the instrumental CMDs relative 
to the RGC centroid for each event.  For three events--KMT-2022-BLG-1551, KMT-2023-BLG-0466, 
and KMT-2025-BLG-0121--the positions of the blend stars are also indicated. Table~\ref{table:five} 
summarizes the source parameters, including the instrumental color and magnitude of the source, 
$(V-I, I)$, and of the RGC centroid, $(V-I, I)_{\rm RGC}$, as well as their de-reddened values, 
$(V-I, I)_0$ and $(V-I, I)_{{\rm RGC},0}$. The table also lists the inferred spectral type and 
angular radius of each source. The sources are found to be main-sequence stars ranging from 
spectral types G to K, with angular radii between 0.34 and 1.14~$\mu$as. Also presented in the 
table are the angular Einstein radius and relative lens-source proper motion. The relative 
lens-source proper motion is calculated by combining the angular Einstein radius and the event 
timescale according to the relation $\mu=\thetae/\te$.

\begin{table*}[t]
\centering
\caption{Physical lens parameters.  \label{table:six}}
\begin{tabular}{lllll}
\hline\hline
\multicolumn{1}{c}{Parameter}               &
\multicolumn{1}{c}{KMT-2020-BLG-0202}       &
\multicolumn{1}{c}{KMT-2022-BLG-1551}       &
\multicolumn{1}{c}{KMT-2023-BLG-0466}       &
\multicolumn{1}{c}{KMT-2025-BLG-0121}       \\
\hline
 $M_{\rm h}$ ($M_\odot$)    &  $0.81^{+0.49}_{-0.36} $  &  $0.54^{+0.31}_{-0.28}$   &  $0.44^{+0.36}_{-0.27}$    &   $0.61^{+0.32}_{-0.32}$      \\   [0.6ex]
 $M_{\rm p}$ ($M_{\rm J}$)  &  $12.28^{+7.37}_{-5.43}$  &  $2.96^{+1.71}_{-1.53}$   &  $3.25^{+2.59}_{-2.03}$    &   $7.77^{+4.08}_{-4.04}$      \\   [0.6ex]
 $\dl$ (kpc)                &  $3.84^{+1.19}_{-1.18} $  &  $6.90^{+0.92}_{-1.33}$   &  $6.80^{+1.14}_{-1.60}$    &   $6.20^{+0.93}_{-1.35}$      \\   [0.6ex]
 $a_\perp$ (AU)             &  $4.17^{+1.29}_{-1.28} $  &  $2.29^{+0.30}_{-0.44}$   &  $2.68^{+0.45}_{-0.63}$    &   $3.05^{+0.46}_{-0.66}$      \\   [0.6ex]
 $p_{\rm disk}$             &  $91\%                 $  &  $34\%                $   &  $31\%                $    &   $43\%                $      \\   [0.6ex]
 $p_{\rm bulge}$            &  $9\%                  $  &  $66\%                $   &  $69\%                $    &   $57\%                $      \\   [0.6ex]
\hline                                                                              
\end{tabular}                                                                      
\end{table*}

\section{Physical lens parameters} \label{sec:five}

We estimated the mass and distance of the lens system by conducting a Bayesian analysis based
on the measured observables of each event. This analysis combines the measured microlensing
observables $\te$ and $\thetae$ with prior information on the Galactic distribution and
kinematics of potential lens and source populations. In this framework, the posterior probability
distribution of the physical parameters is expressed as
\begin{equation}
P(M, \dl ~|~ \te, \thetae) \propto P(\te, \thetae ~|~ M, \dl) P(M, \dl),
\label{eq3}
\end{equation}
\hskip-4pt
where the likelihood, $P(\te, \thetae ~|~ M, \dl)$, quantifies how well a model with given $M$
and $\dl$ reproduces the observed $\te$ and $\thetae$, and the prior, $P(M, \dl)$, represents
the expected distributions of lenses and sources in the Galaxy.

The priors were derived from a Galactic model that includes the density and velocity distributions
of disk and bulge stars, as well as a mass function for the lens population. In the analysis, we
adopted the Galactic model of \citet{Jung2021} and employed the lens mass function proposed
by \citet{Jung2022}. Using these priors, a large number of lens-source pairs were generated via
Monte Carlo simulation. For each simulated event, the observables $\te$ and $\thetae$ were
computed, and the likelihood of each event was evaluated as
\begin{equation}
L \propto \exp \left( -{\chi^2 \over 2}\right), 
\chi^2 = 
\frac{(t_{\rm E}-t_{\rm E,obs})^{2}}{\sigma_{t_{\rm E}}^2} +
\frac{(\theta_{\rm E}-\theta_{\rm E,obs})^{2}}{\sigma_{\theta_{\rm E}}^2}.
\label{eq4}
\end{equation}
\hskip-4pt
Here, $(t_{\rm E,obs}, \theta_{\rm E,obs})$ denote the observed values of the lensing observables,
and $(\sigma_{t_{\rm E}}, \sigma_{\theta_{\rm E}})$ represent their measurement uncertainties.

\begin{figure}[t]
\includegraphics[width=\columnwidth]{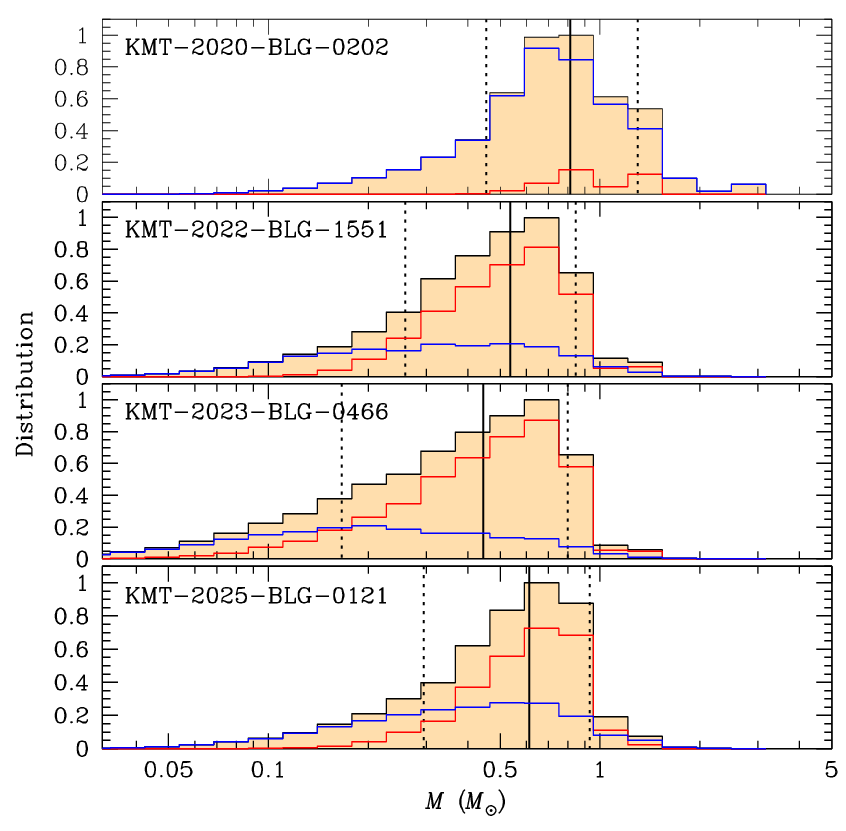}
\caption{
Posterior probability distributions of the lens mass. In each panel, the probability contributions
from the disk and bulge lens populations are shown by blue and red curves, respectively, while the 
total distribution is indicated by a black curve.  The solid vertical line represents the median 
of the distribution, while the two dotted vertical lines indicate the 1$\sigma$ range.
}
\label{fig:eight}
\end{figure}

\begin{figure}[t]
\includegraphics[width=\columnwidth]{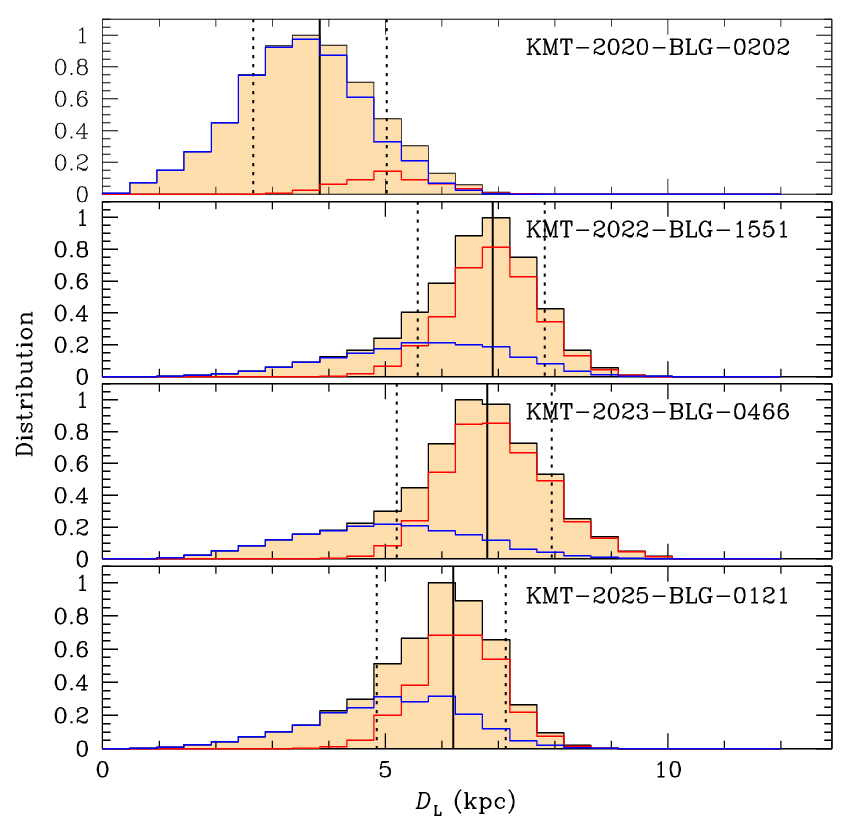}
\caption{
Posterior probability distributions of the distance to the lens.
}
\label{fig:nine}
\end{figure}

The posterior probability distributions of the events are shown in Figure~\ref{fig:eight} for the 
lens mass and in Figure~\ref{fig:nine} for the lens distance. The derived physical parameters 
of the lens system are summarized in Table~\ref{table:six}, where $M_{\rm p}$ and $M_{\rm h}$ 
denote the masses of the planet and its host, respectively, and $a_\perp$ represents the projected 
separation between them.  The median of each posterior distribution was adopted as the representative 
value, and the lower and upper uncertainties correspond to the 16th and 84th percentiles, respectively.  
Also listed are the probabilities that the lens resides in the Galactic disk ($p_{\rm disk}$) or in 
the bulge ($p_{\rm bulge}$).

It is found that the companions to the lenses are super-Jupiters, with masses exceeding that 
of Jupiter but below the deuterium-burning threshold of $\sim 13~M_{\rm J}$ \citep{Burrows1997, 
Spiegel2011}.  For KMT-2020-BLG-0202L, the upper limit of the companion mass exceeds this 
threshold, although the median value lies below it, implying that the object may be a brown 
dwarf. In all cases, the host stars have masses lower than that of the Sun.

The location of the snow line is expected to scale with the host mass as $a_{\rm sl} \sim 
2.7~{\rm AU}\,(M/M_\odot)$ \citep{Kennedy2008}.  Given that the measured $a_\perp$ values 
represent projected separations, all of the detected planets lie well beyond the snow line 
of their respective hosts.  Detection of such cold giant planets by other methods, such as 
the radial-velocity or transit techniques, is inherently challenging owing to their long 
orbital periods and the low geometric probability of transit alignment.  The detections 
reported here therefore highlight the unique capability of the microlensing method to probe 
the population of cold, massive planets beyond the snow line.

The inferred lens locations differ from event to event. For KMT-2020-BLG-0202, the lens 
is most likely in the Galactic disk, with a probability of $p_{\rm disk}=91\%$. By contrast, 
for KMT-2022-BLG-1551 and KMT-2023-BLG-0466, the lenses are about twice as likely to reside 
in the bulge as in the disk. For KMT-2025-BLG-0121, the probabilities of disk and bulge 
membership are approximately equal.

\section{Summary and conclusion} \label{sec:six}

We have analyzed four microlensing events--KMT-2020-BLG-0202, KMT-2022-BLG-1551,
KMT-2023-BLG-0466, and KMT-2025-BLG-0121--that exhibit extended and complex anomalies in
their light curves. These events were identified through a systematic reanalysis of KMTNet data
aimed at recovering planetary signals that deviate from the typical short-term anomaly morphology.
Detailed light-curve modeling shows that the anomalies in all four cases are produced by planetary
companions orbiting low-mass stellar hosts.

The planetary systems have mass ratios in the range $q \sim (5$--$14)\times 10^{-3}$ and
Einstein timescales of $t_{\rm E} \sim 20$--$45$ days. Bayesian analyses based on Galactic
models indicate that the companions are super-Jupiters, with masses of a few to $\sim$ 10
$M_{\rm J}$, orbiting sub-solar-mass host stars. All planets lie well beyond the snow line of their
respective hosts, placing them firmly in the cold-giant regime. This population provides valuable
insight into the occurrence of massive planets at wide separations, a region of parameter space
largely inaccessible to other detection techniques.

These results demonstrate that extended and complex anomaly patterns--often missed by automated
detection algorithms--can contain planetary signals. Failure to recover such planets would bias
demographic inferences, leading to underestimated occurrence rates and distortions in the inferred
planet-to-host mass-ratio function, particularly for high-$q$ planets. Systematic reanalysis of
microlensing survey data is therefore essential for constructing a more complete and less biased
census of exoplanets in the Galaxy. While no approach can guarantee complete recovery of all
planetary signals, sustained efforts along these lines will reduce selection biases in demographic
studies and thereby improve our understanding of the formation and distribution of cold giant
planets.

\citet{Zang2025} reported the mass-ratio distribution of planets based on KMTNet detections
from the 2016--2019 seasons. Their statistical sample includes all planets detected in 2018 and
2019, whereas for earlier seasons only planets with mass ratios of order $10^{-4}$ and below
were included because the detection efficiency for higher-$q$ planets had not yet been fully
quantified. Consequently, of the three planets presented in Paper I, only the 2019-season event
(OGLE-2019-BLG-0299Lb) is included in their analysis, while the two 2017-season events
(KMT-2017-BLG-2509Lb and OGLE-2017-BLG-1099Lb) are not. The KMTNet collaboration plans to
revisit this analysis with an expanded sample that incorporates planets reported since the 2020
season, including those presented here and those omitted from the initial study.

A further consideration for detection efficiency, and hence inferred demographics, is binary-lens
degeneracy, although the four planets reported here are not affected by this issue given their
well-constrained solutions. \citet{Shang2025} showed that 2L1S degeneracies reduce survey
sensitivity by $\sim$ 5\%--10\%, with the impact increasing toward higher mass ratios. If neglected,
this leads to underestimated planet occurrence rates and a flatter inferred mass-ratio function.
Future demographic analyses should therefore incorporate degeneracy-aware sensitivity estimates to
mitigate such systematic biases as microlensing samples continue to grow.

\begin{acknowledgments}
C.H. was supported by the National Research Foundation of Korea (NRF) grant funded by the 
Korea government (MSIT: RS-2025-21073000).
This research was supported by the Korea Astronomy and Space Science Institute under the R\&D 
program (Project No. 2025-1-830-05) supervised by the Ministry of Science and ICT.
This research has made use of the KMTNet system operated by the Korea Astronomy and Space Science 
Institute (KASI) at three host sites of CTIO in Chile, SAAO in South Africa, and SSO in Australia. 
Data transfer from the host site to KASI was supported by the Korea Research Environment Open NETwork 
(KREONET). 
%
H.Y. and W.Z. acknowledge support by the National Natural Science Foundation of China 
(Grant No. 12133005). H.Y. acknowledge support by the China Postdoctoral Science Foundation 
(No. 2024M762938).
\end{acknowledgments}



\bibliographystyle{aasjournal}
\bibliography{pasp_refs}

\end{document}